\documentclass[twocolumn]{openjournal}
\usepackage{natbib}
\usepackage{graphicx,amsmath,amssymb,amstext}
\usepackage{amsbsy,amsfonts,amsthm,color}
\usepackage[colorlinks,linkcolor=blue,citecolor=blue,urlcolor=blue ]{hyperref}
\usepackage[utf8]{inputenc}
\usepackage{float}
\usepackage{graphicx}

\usepackage{subcaption}

\defcitealias{Prole2023}{LP23}
\defcitealias{Schauer2021}{AS21}

\begin{document}


\title{Primordial black holes in cosmological simulations: growth prospects for supermassive black holes}

\author{Lewis R. Prole$^{* \ 1}$}
\author{John A. Regan$^{1}$}
\author{Daxal Mehta$^{1}$}
\author{Peter Coles$^{1}$}
\author{Pratika Dayal$^{2}$}
\email{$^*$email: lewis.prole@mu.ie}
\affiliation{Centre for Astrophysics and Space Sciences Maynooth, Department of Theoretical Physics, Maynooth University, Maynooth, Ireland.}
\affiliation{Kapteyn Astronomical Institute, University of Groningen, PO Box 800, 9700 AV Groningen, The Netherland}

\begin{abstract}
It has long been suggested that a fraction of the dark matter in the Universe could exist in the form of  primordial black holes (PBHs) that have existed since the radiation dominated era. Recent studies have suggested that these PBHs may be the progenitors to the population of high-redshift, supermassive black holes (SMBHs) observed since the launch of JWST. For the first time, we have included PBHs in cosmological simulations, to test whether PBHs can sink to the center of collapsing halos, locate dense gaseous regions and experience significant growth. We tested PBH-to-DM mass ratios of $f_{\rm PBH}$ = $10^{-4}$ and $10^{-3}$, with an initial PBH mass of 1000 M$_\odot$, as inspired by recent observational constraints. We find that at $f_{\rm PBH} = 10^{-3}$, a number of PBHs were able to embed themselves in dense gas and grow to $10^{4}$-$10^{5}$ M$_\odot$ by $z=20$. These intermediate black holes (IMBHs) are possible progenitors to the highest redshift SMBH observations, such as GNZ-11 ($10^{6}$ M$_\odot$ by $z=10$), outperforming light seed black hole (BH) growth seen in recent simulations without the need to invoke heavy seeding prescriptions. On the other hand, $f_{\rm PBH} = 10^{-4}$ resulted in no significant BH growth, emphasizing that the ability of PBHs to act as SMBH seeds is sensitive to the true value of $f_{\rm PBH}$ in the Universe, and showing that the $f_{\rm PBH} =10^{-4}-10^{-3}$ boundary marks the threshold above which SMBH seeding via 1000 M$_\odot$ PBHs becomes effective. This is the first step towards building a realistic PBH framework in cosmological simulations.
\end{abstract}

\keywords{}

\maketitle

\section{Introduction} 
\label{sec:intro}
Supermassive black holes (SMBHs) have long been observed at redshifts $z = 6-7$ with masses up to $10^{10}$ M$_\odot$ (see \citealt{Inayoshi2020} for a review). With the recent launch of the James Webb Space Telescope (JWST), we now have observations of SMBHs with masses $10^7 - 10^8$ M$_\odot$ existing as early as $z \sim 9$. These objects have been detected both directly in the CEERS survey \citep{Larson2023} and via gravitational lensed sources in the UNCOVER survey \citep{Kovacs2024,Bogdan2024,Kovacs2024}. The discovery of such massive objects in the early Universe challenges black hole (BH) growth models in a standard $\Lambda$CDM Universe. Moreover, lower mass 10$^6$ M$_\odot$ SMBHs have been observed out to $z \sim 11$ with the JADES survey \citep{Maiolino2023a,Maiolino2024}. These objects have BH to stellar mass ratios as much as 1000 times higher than the local relation predicts, further straining our understanding of high redshift galaxy evolution.

Adding to the confusion is the observation of a population of compact, red objects at $z = 4 - 10$ known as little red dots \citep[LRDs;][]{Matthee2023,Matthee2024, Harikane2023, Kocevski2023}. These objects are accompanied by broad line emission, usually indicative of the presence of a massive black holes (MBH). Inferred bolometric luminosities suggest these objects house SMBHs of mass $10^7 - 10^8$ M$_\odot$ \citep{Greene2024}. If true, this implies that the MBH number densities are higher than previously expected (see also \citealt{Akins2024}).

Traditional attempts to account for the existence of high-redshift SMBHs typically rely on either light seed BHs (remnants of Population III stars) accreting at super-Eddington rates (e.g. \citealt{Pezzulli2016,AlisterSeguel2020,Mehta2024}) or so-called heavy BH seeding scenarios, where massive star or direct collapse BH formation is triggered by external radiation fields \citep{Ardaneh2018,Patrick2023,Prole2024a}. However, the radiation flux required ($>10^4 J_{21}$: \citealt{Sugimura2014,Glover2015,Agarwal2015,Agarwal2016}) is much higher than the average flux a halo would receive at high redshift \citep{Ahn2009,Trenti2009,Wise2012,Agarwal2012,Skinner2020}, rendering the scenario very rare (e.g. \citealt{Dijkstra2008,OBrennan2024}). Alternatively, DM steaming velocities \citep{Greif2011b,Schauer2017,Schauer2021}, rapid halo assembly \citep{Wise2019,Regan2020a,Mone2025} or large-scale colliding flows \citep{Latif2022, Prole2024} have been shown to boost stellar/BH masses. 

An answer to the SMBH problem without the need for heavy seeding prescriptions may be found in a theoretical BH formation channel commonly overlooked in analytical work and simulations. It is speculated that the initial density perturbations created from quantum fluctuations could have
collapsed into BHs during the first second after the big bang \citep{Carr1974}. This population of primordial black holes (PBHs) would have existed long before the formation of the first stars or the collapse of the first cosmological halos (see e.g. \citealt{Dayal2024,Zhang2025}). The idea is supported by observations of BHs in mass bands not covered by astrophysical processes, with masses either high enough to fall into the pair-instability supernovae (SNe) band \citep{LIGOScientificCollaborationandVirgoCollaboration2020} or below the predicted minimum BH mass \citep{Abac2024}. Additionally, PBHs are invoked to explain low chemical enrichment observed in systems containing massive BHs \citep{Maiolino2025}. However, even if PBHs do exist, their growth prospects are far from certain, as accretion of low density gas onto BHs is inefficient (e.g. \citealt{Edgar2004}). The ability to accrete up to SMBH masses by the time of JWST observations therefore relies on PBHs finding and sinking into the center of dark matter (DM) halos, where baryonic gas collapses to higher densities prior to star formation.

As PBHs formed before big bang nucleosynthesis (BBN), they are not subject to the BBN derived constraint that baryons can have at most 5$\%$ of the
critical density \citep{Cyburt2003}. PBHs are therefore an ideal candidate to explain the existence of DM, by accounting for all or a percentage of the unseen mass attributed to DM (\citealt{Zwicky1933}, see \citealt{Arbey2021} for a recent review).

Observational constraints have been placed on the possible masses of PBHs and fraction of DM that can exist as PBHs today, $f_{\rm PBH}$. While PBHs lighter than 10$^{-19}$ M$_\odot$ are expected to have evaporated via Hawking radiation by the present day \citep{Hawking1975, Auffinger2023}, current constraints allow very low mass PBHs ($10^{-16}-10^{-12}$ M$_\odot$) to theoretically constitute 100$\%$ of the DM \citep{Raidal2019}. However, the mass resolution required to test this regime makes it unfeasible with current cosmological simulation codes, so the work presented here will focus on higher mass PBHs. Broadly, we know from lensing of type Ia SNe that the abundance of all compact objects in the Universe above the mass of 0.01 M$_\odot$ is constrained to be less than 35$\%$ of the total mass, or $f_{\rm PBH} \lesssim 0.5$  \citep{Zumalacarregui2018}. Additionally, 
$f_{\rm PBH}$ is limited at $\lesssim 10^{-2}$ in the range of $0.1-1$ M$_\odot$ by microlensing in the Optical
Gravitational Lensing Experiment (OGLE) \citep{Mroz2024}. The mass range of $0.5 - 300$ M$_\odot$ is currently most constrained by data from the third observational run of LIGO-Virgo-KAGRA, which limits $f_{\rm PBH}$ to 10$^{-3}$ at low masses and decreases to 10$^{-4}$ by 100 M$_\odot$ \citep{Andres-Carcasona2024}. The high end of PBH masses is constrained by cosmic microwave background (CMB) observations, as PBH accretion feedback from PBHs is expected to effect the thermal history of the universe. Although it is generally agreed that the upper limits of $f_{\rm PBH}$ fall off steeply with the assumed PBH mass, estimates assuming disk accretion onto PBHs give stringent limits of $f_{\rm PBH} \lesssim 10^{-3} - 10^{-4}$ for $10^3$ M$_\odot$ PBHs, while spherical accretion allows a much higher $f_{\rm PBH}\lesssim 0.05$ \citep{Piga2022,Agius2024}. However, constraints from dwarf galaxy heating also limit $f_{\rm PBH}\lesssim 10^{-3}$ for $10^3$ M$_\odot$ PBHs \citep{Lu2021}.

Large-volume cosmological simulations including aspects related to PBHs are scarce, but the investigations that have been performed so far give interesting initial insights into the behavior of PBHs in a cosmological context. Purely N-body simulations of PBHs \citep{Trashorras2021} show that PBHs form stable 100 kpc clusters made up of multiple 1 kpc sub-clusters that survive until the present day, with roughly 10$\%$ of PBHs forming binaries and the most massive PBHs resulting from major mergers tending to segregate themselves within the 100 pc cluster core. Cosmological N-body simulations of PBHs and DM \citep{Tkachev2020} show that the merger rate of PBHs matches the merger rate of binary BHs estimated by the LIGO collaboration \citep{Abbott2016} for $f_{\rm PBH}=10^{-4}$, a result which is seemingly independent of the chosen PBH mass. 

By modeling PBHs using a semi-analytic model alongside a hydrodynamical simulation, \cite{Liu2022} showed that PBHs act to accelerate structure formation (giving BH seeds more time to accrete) and shift star formation to more massive halos. Likewise, by making changes to the initial mass power spectrum to account for the presence of PBHs, \cite{Colazo2024}  reproduced similar levels of high redshift galaxy abundances as found with JWST observations.

Analytical work suggests that PBHs not only provide the seeds for SMBHs \citep{Ziparo2025}, but that massive PBHs can seed entire halos \citep{Dayal2024,Zhang2025}, resulting in galaxies with elevated BH to stellar mass ratios. Singular PBHs altering structure formation in this manner is often referred to as the `Coulomb effect'.

Despite these first steps towards understanding the role PBHs may play throughout cosmic history, a population of PBHs constituting a fraction of DM (as constrained by observations) has never been directly included in hydrodynamical simulations, collectively affecting their local environment in real time by taking into account the properties and diversity of their surroundings and the clustering behavior of PBHs (often referred to as the `Poisson effect'). Only once PBHs are modeled self-consistently in this way will simulations allow us to see definitively what effects they have on structure formation and SMBH growth. In this paper, our objective is to test whether a population of PBHs, modeled as sink particles, with initial masses of 1000 M$_\odot$ and constrained mass fractions $f_{\rm PBH} = 10^{-4} - 10^{-3}$ of DM, are capable of growing prior to the earliest SMBH observations at $z = 10$. As SMBH formation via heavy seeding remains unproven and far from certain, the ability of PBHs to sink into halos and accrete will make them a viable alternative SMBH formation pathway.

The structure of the paper is as follows; in Section \ref{sec:method} we present the numerical method, including the initial conditions, DM to PBH particle conversion, zoom simulations, resolution, the PBH accretion scheme and PBH mergers. In Section \ref{sec:results} we present the results of the simulations before discussing future steps and caveats in Section \ref{sec:caveats} and concluding in Section \ref{sec:conclusions}. In Appendix \ref{sec:appendix}, we present an additional simulation exploring the effects of DM particle mass resolution and mass disparity with PBHs.
\\
\\
\section{Numerical method}
\label{sec:method}

\subsection{Initial conditions and PBH seeding}
The simulations we present here were performed in 2 stages; the parent simulation was initialized using the initial conditions generator MUSIC \citep{Hahn2011}, giving a parent co-moving simulation box of side length $40 \: h^{-1} \: {\rm Mpc}$ at $z=127$. We used the $\Lambda$CDM cosmology parameters $h=0.6774$, $\Omega_0 = 0.3089$, $\Omega_{\rm b} = 0.04864$, $\Omega_\Lambda = 0.6911$, $n = 0.96$ and
$\sigma_8 = 0.8159$ \citep{Planck-Collaboration2020}.  The simulations were initialized with a DM distribution using the transfer functions of \cite{Eisenstein1998}. The gas distribution was set to initially follow the dark matter. We applied a Jeans refinement criterion such that the Jeans length of the gas was always resolved with at least 4 grid cells. After running the simulation to $z = 20$, we identified the largest halo in the box using a Friends-of-Friends (FOF) algorithm and used it as the central coordinate for the second stage, zoom-in simulations.

The simulations were then re-initialized at $z=127$ with a zoom region centralized on the most massive halo from the initial simulation. The resimulation had 14 levels of refinement, with the most refined grid level spanning $0.5 \: h^{-1}$ co-moving $\: {\rm Mpc}$ and containing 8741816 DM particles, with a mass resolution of 1600  M$_\odot$, and an additional 8741816 gas cells. Within the zoom region, we randomly sampled DM particles to be converted into PBH particles to give the desired PBH to DM mass ratio $f_{\rm PBH}$. We repeated the simulation to test values $f_{\rm PBH}$ = $10^{-4}$ and $10^{-3}$ and ran the simulations to $z=20$. As the DM particles were initially spread roughly equally across the 640 processors used to perform the simulations, we calculated the number of DM particles to be converted by applying $f_{\rm PBH}$ to the number of particles on that processor, which is not guaranteed to be an integer number. As such, the total number of DM particles converted to PBHs across all processors N$_{\rm PBH}$ does not exactly match the desired $f_{\rm PBH}$. We give the target and real values of $f_{\rm PBH}$ and N$_{\rm PBH}$ in Table \ref{table:PBH}.

  \begin{table}[!ht]
    \caption[Table]{ Number of PBHs in each simulation - target values of $f_{\rm PBH}$ and number of PBHs N$_{\rm PBH}$ versus actual values of $f_{\rm PBH}$ and N$_{\rm PBH}$ produced by the random sampling across multiple processors at $z=127$.}\label{table:PBH}
    \centering
      \small
      \begin{tabular}{cccc}
                \hline
                 $f_{target}$ & N$_{target}$ & $f_{real}$ & N$_{real}$ \\
                \hline
                 10$^{-4}$ & 874 & $0.810 \times 10^{-4}$ & 707 \\
                 10$^{-3}$ & 8726 & $0.998 \times 10^{-3}$ & 8726 \\
                \hline
      \end{tabular}
  \end{table}

\subsection{Resolution and PBH accretion}
We allowed Jeans refinement of the gas down to scales of 5 co-moving pc ($\sim$ 0.3 physical pc at $z = 20$). We therefore set the PBHs accretion radius, $R_{\rm acc}$ to 5 times the minimum cell length, $R_{\rm acc}$ = 25 co-moving pc. 

We model accretion onto PBHs as Bondi accretion. The Bondi radius is given by \citep{Bondi1952} 
\begin{equation}
        R_{\rm Bondi} = \frac{G M_{\rm sink}}{v_{\infty}^2+c_{\infty}^2},
        \label{eq:bondi_radius}
\end{equation}
where $M_{\rm sink}$ is the mass of the sink particle, $v_{\infty}$ is the mass weighted speed of the gas within the accretion radius relative to the BH and $c_{\infty}$ is the sound speed in the region. In instances where the accretion radius is larger than $R_{\rm Bondi}$, we instead opt to use $R_{\rm Bondi}$ for as the accretion zone to drain mass from gas cells.

The accretion rate onto the BH is then calculated using the usual Bondi formula
\begin{equation}
    \dot{M}_{\rm Bondi} = 4 \pi \rho_{\infty} R_{\rm Bondi}^2 ((1.12 c_{\infty})^2 + v_{\infty}^2)^{1/2},
    \label{eq:bondi-hoyle-acc}
\end{equation}
where $\rho_{\infty}$ is the weighted density inside the accretion sphere. As in \cite{Krumholz2004}, we weight the density by assigning weights to all cells within a kernel radius, $r_K$, given by 
\begin{equation}
\rm{r_K} = \left\{ \begin{array}{lcr}
  \Delta x_{\rm min}/4 & &R_{\rm Bondi} <  \Delta x_{\rm min}/4\\
  R_{\rm Bondi}  & & \ \ \ \ \Delta x/4 \le R_{\rm Bondi} \le R_{\rm acc}\\
  R_{\rm acc} && R_{\rm Bondi} > R_{\rm acc},
\end{array} \right.
\end{equation}
where $\Delta x_{\rm min}$ is the current minimum cell volume. $\rm{r_K}$ is used to assign a weight to every cell within $R_{\rm acc}$ using
\begin{equation}
    W \propto \rm{exp(-r^2/r_{K}^2)},
\label{eq:weights}
\end{equation}
where $r$ is the distance from the cell to the accreting BH. $\rho_{\infty}$ is then calculated as
\begin{equation}
     \rho_{\infty} = \bar{\rho}  W,
\end{equation}
where $\bar{\rho}$ is the mass weighted mean density within the accretion sphere.

\subsubsection{Vorticity adjustment}
Following on from \cite{Krumholz2006}, we adjust the accretion rate based on the vorticity $\omega$ of the surrounding gas, given by
\begin{equation}
    \omega = | \nabla \times \vec{v} |,
\end{equation}
with the dimensionless vorticity $\omega_*$ given by
\begin{equation}
\omega_* =  \omega  \frac{R_{\rm Bondi}}{c_{\infty}}.
\end{equation}
We introduce a damping factor $f(\omega)$ defined as
\begin{equation}
    f_{w} = \frac{1}{1 + \omega_*^{0.9}},
\end{equation}
and calculate the accretion rate in a turbulent medium according to
\begin{equation}
    \dot{M}_{\omega} = 4  \pi  \rho_{\infty}  R_{\rm Bondi}^2  c_{\infty}  (0.34  f_{\omega_*}).
\end{equation}
The total accretion rate onto the MBH particle is
\begin{equation}
    \dot{M} = (\dot{M}_{\rm Bondi}^{-2} + \dot{M}_{\omega}^{-2})^{-0.5},
\end{equation}

For a given time step, $t_h$, the mass of the MBH particle increases by $t_h \dot{M}$, which is removed from cells within $R_{\rm acc}$ using the weighting scheme calculated in Equation \ref{eq:weights}, adjusting the velocities of both the gas cells and the BH to conserve linear momentum. The BH position is then shifted to the center of mass of a system comprised of the the BH and the accreted mass contributions from each cell.

\subsection{PBH mergers}
We allow PBH particles to merge based on the treatment originally implemented in \cite{Prole2022}. PBH particles are merged if:
   \begin{itemize}
      \item They lie within each other’s accretion radius.
      \item They are moving towards each other.
      \item Their relative acceleration is $<$0.
      \item They are gravitationally bound to each other.
   \end{itemize}
As PBH particles carry no thermal data, the last criteria simply require
that their gravitational potential exceeds the kinetic energy of
the system. When these criteria are met, the larger of the particles gains
the mass and linear momentum of a smaller particle, and its position
is shifted to the centre of mass of the system. 

\subsection{Lyman-Werner radiation field}
We account for the presence of star formation indirectly in our simulations via an evolving a Lyman-Werner (LW) background radiation field, which is used within the chemical network to modify the rates of chemical reactions. We extrapolate the background LW intensity from the reference simulation of \cite{Qin2020} as presented in their Figure 5. In units of J$_{21}$ = 10$^{-21}$ erg s$^{-1}$ Hz$^{-1}$ sr$^{-1}$ cm$^{-2}$, the field begins at $z \sim 30$ with a value of $\sim$ 0.004 J$_{21}$ and climbs to $\sim$ 2 J$_{21}$ by $z = 10$.

\begin{figure}
\centering
         \hbox{\hspace{0cm} \includegraphics[width=1\linewidth]{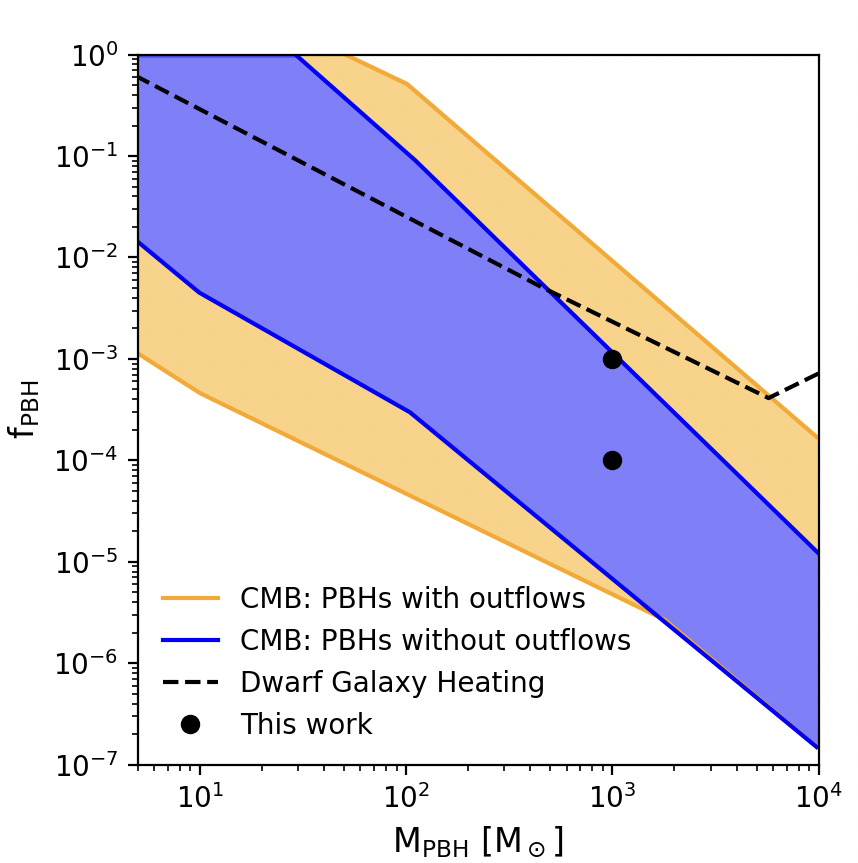}}
    \caption{Values of $f_{\rm PBH}$ adopted in this work with  relation to the constrained upper limits of $f_{\rm PBH}$ as a function of PBH mass. The shaded areas represent the regions of uncertainty in the limit from CMB measurements, considering PBHs with (orange) and without (blue) outflows \citep{Piga2022}. The the black dashed line gives constraints from dwarf galaxy heating \citep{Lu2021}.}
    \label{fig:constraints}
\end{figure}

\begin{figure*}
\centering
         \hbox{\hspace{0cm} \includegraphics[width=1\linewidth]{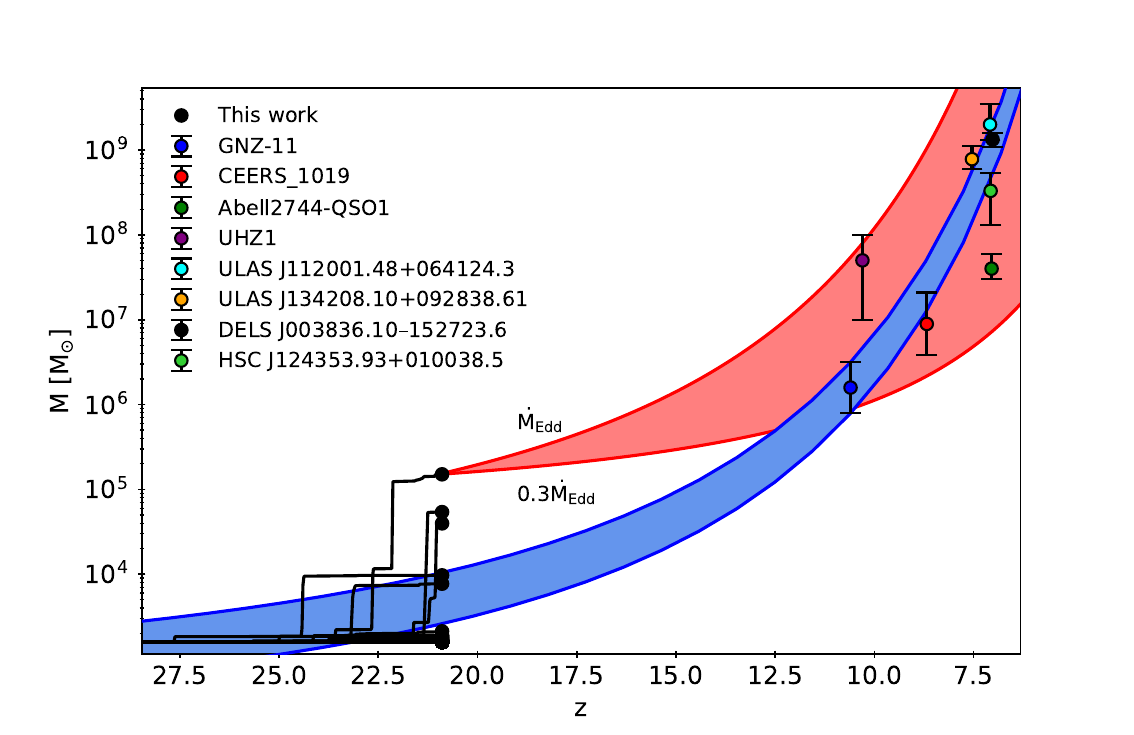}}
    \caption{Growth of PBHs as a function of redshift for our $f_{\rm PBH}$ = 10$^{-3}$ run. To guide the eye, the blue shaded region represents the Eddington limited growth required to achieve JWST observed 10$^6$ M$_\odot$ SMBH GNZ-11 \citep{Maiolino2023} at $z \sim 10$. We also show SMBH observations from CEERS$\_$1019 \citep{Larson2023}, Abell2744-QSO1 \citep{Furtak2024}, UHZ1 \citep{Bogdan2024}, ULAS J112001.48+064124.3 \citep{Mortlock2011}, ULAS J134208.10+092838.61 \citep{Banados2018}, DELS J003836.10-152723.6 \citep{Wang2018} and HSC J124353.93+010038.5 \citep{Matsuoka2019a}. The red shaded region shows that the growth required for our largest PBH to match these observations lies between 0.3-1 times the Eddington limit.}
    \label{fig:growth}
\end{figure*}

\begin{figure}
    \centering
    \begin{subfigure}{\linewidth}
    \centering
    \includegraphics[width=0.9\linewidth]{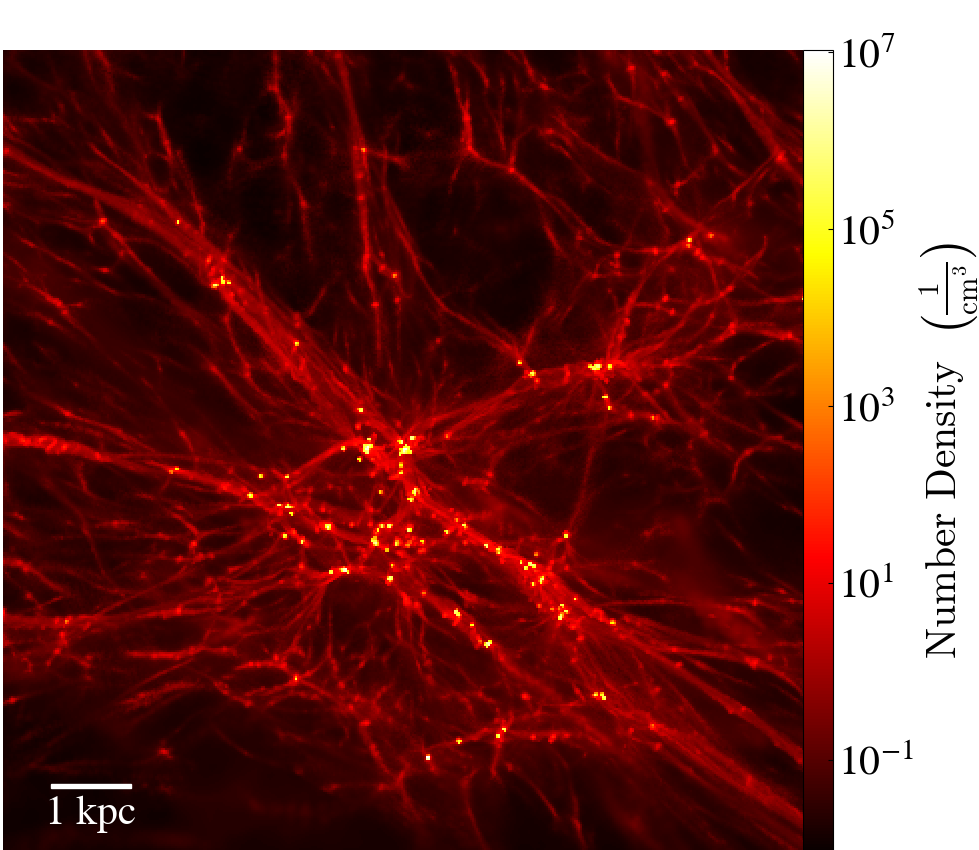}
    \end{subfigure}
    \begin{subfigure}{\linewidth}
    \centering
    \includegraphics[width=0.9\linewidth]{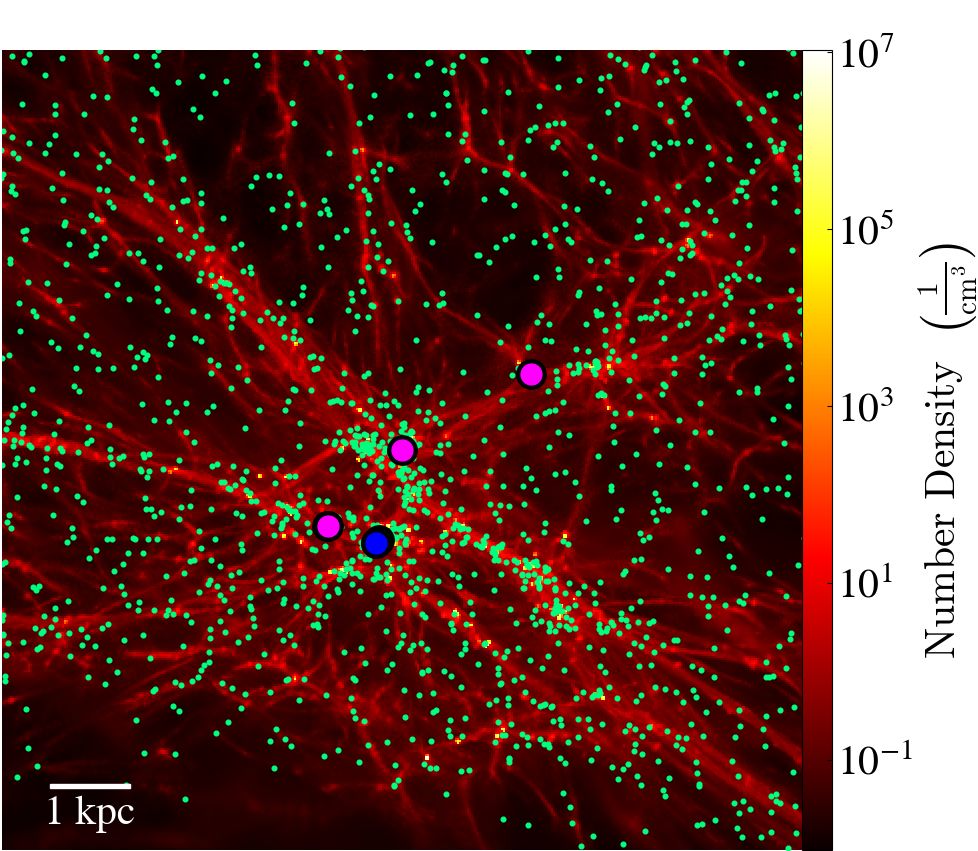}
    \end{subfigure}
   \begin{subfigure}{\linewidth}
    \centering
    \includegraphics[width=0.9\linewidth]{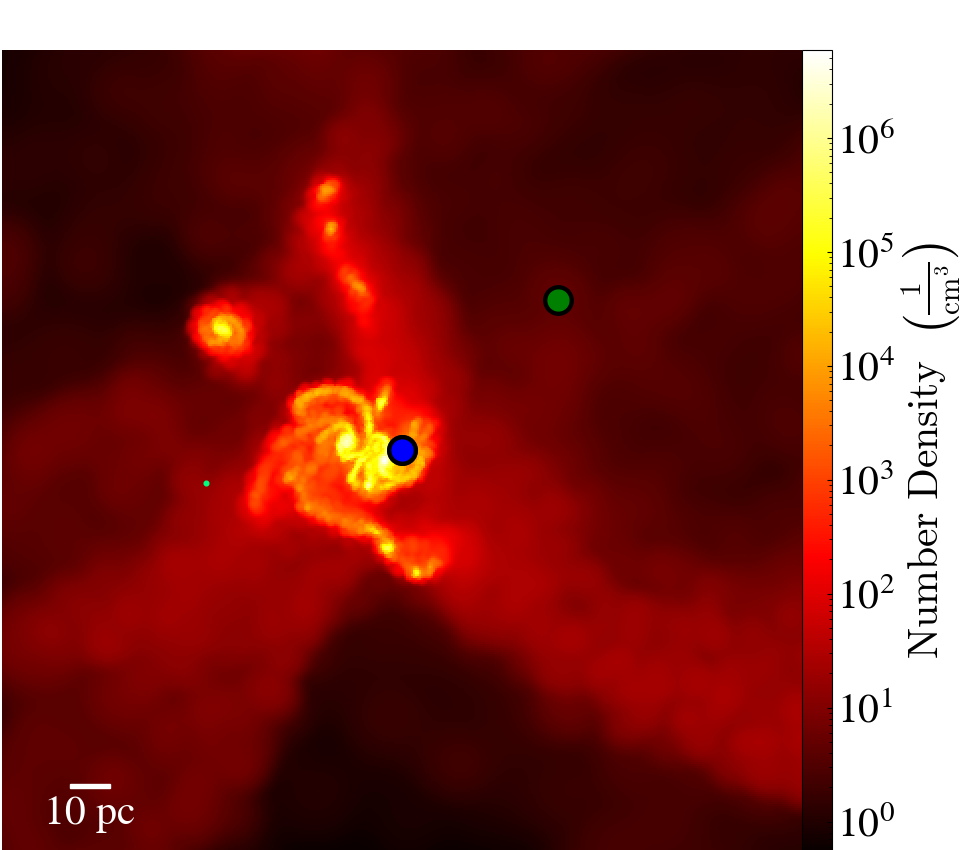}
    \end{subfigure}
    \caption{Density projections of the $f_{\rm PBH} = 10^{-3}$ simulation. Top - 10 kpc region centered on the growing PBHs, shown at the end of the simulation ($z \sim 21$). Middle - the same region with all PBHs overplotted. We show PBHs with no significant growth (more than 50$\%$ of their initial mass) as small lime green dots, while growing PBHs are shown as larger magenta circles. We also show the most massive PBH as a blue circle, with its nearest neighboring PBH as a green circle to match the colorscheme of Figure \ref{fig:environment}. Bottom - a 200 pc zoom-in region around the most massive PBH. }
    \label{fig:vertical_subplots}
\end{figure}

\begin{figure}
\centering
          \includegraphics[width=1\linewidth]{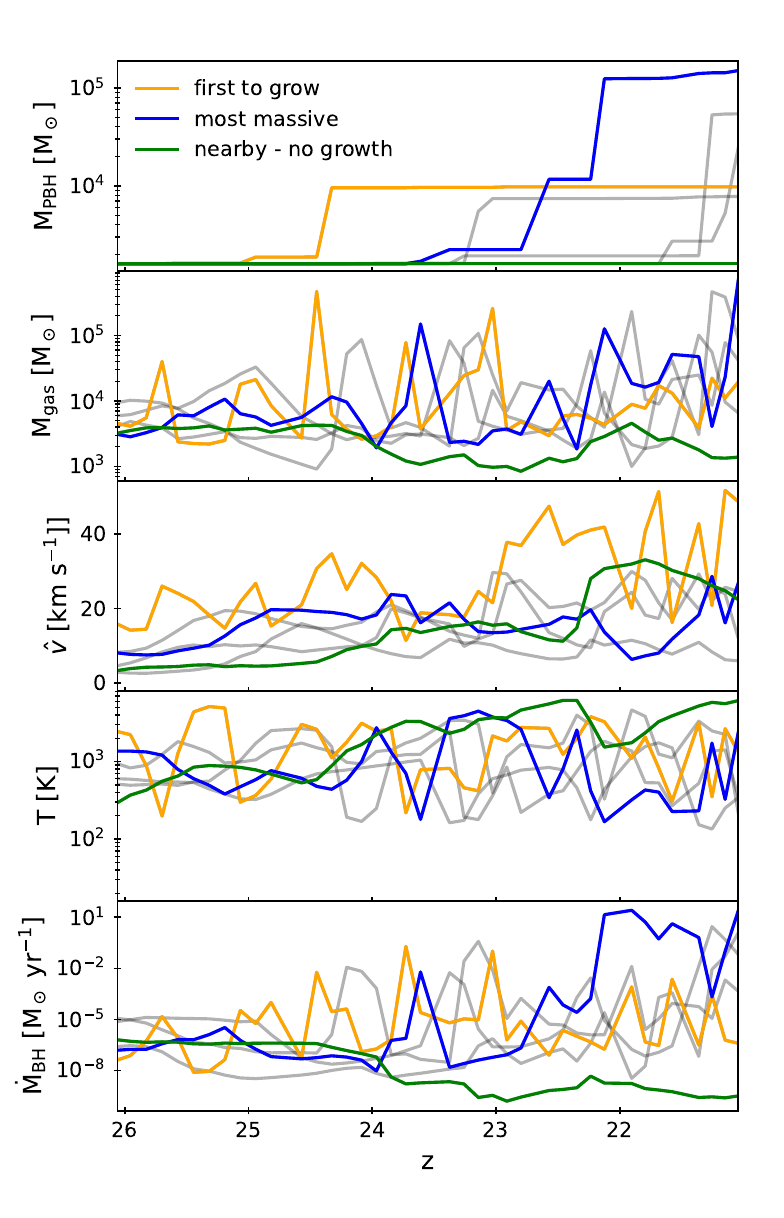}
    \caption{Redshift evolution of the mass averaged quantities in the environment surrounding growing PBHs. From top to bottom - the mass of the PBH, the total gas mass within a 100 co-moving pc region around the PBHs, the average relative velocity between the surrounding gas and the PBH, the average temperature of the gas and the estimated Bondi-Hoyle accretion rate. We show these values for all PBHs that grew beyond 10$^{4}$ M$_\odot$ as black transparent lines, and highlight special cases; the first PBH to start growing (orange), the most massive PBH by the end of the simulation (blue), the most massive PBH's nearest neighboring PBH (green) as an example of a  PBH that did not grow at all.}
    \label{fig:environment}
\end{figure}

\begin{figure*}
\centering
          \includegraphics[width=1\linewidth]{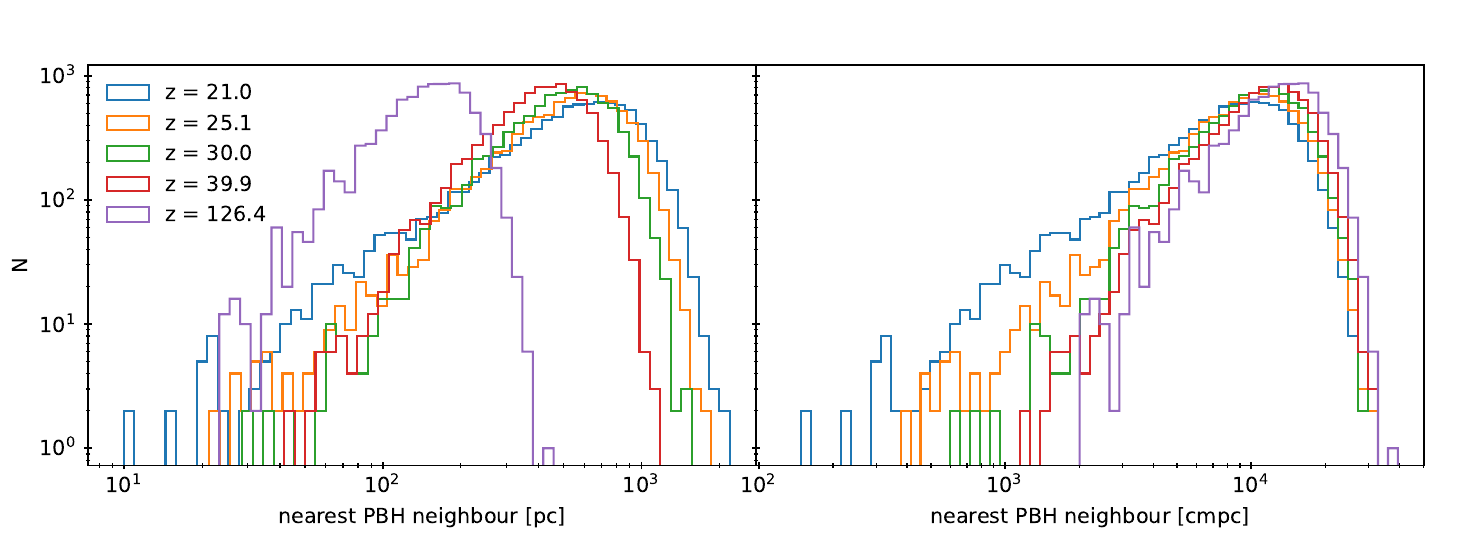}
    \caption{Clustering behavior of PBHs shown through histograms of the distance to the nearest neighboring PBH. We show a snapshot soon after the initial conditions at $z=126$, and 4 snapshots equally spaced in terms of cosmic scale factor $a=1/(1+z)$, from $z=40$ to the end of the simulation at $z=21$. Left - distribution as a function of physical distance. Right - distribution of a function of co-moving distance.}
    \label{fig:spacing}
\end{figure*}


\section{Results}
\label{sec:results}
We ran the simulations until $z=20$, as this is a point where typically the first stars and astrophysical BHs have already started forming in large-scale cosmological simulations \citep{Bellovary2019,Schauer2021, Regan2023a,Bhowmick2024}. As such, PBH growth at later times has little advantage over conventional heavy seeding channels.

Zero PBHs in the $f_{\rm PBH}=10^{-4}$ run experienced significant growth, which we define as at least a doubling of their initial mass. The significance of this result is made clear by Figure \ref{fig:constraints}, which shows where our chosen values of $f_{\rm PBH}$ fit within existing observational constraints. There is an uncertainty in the true upper limit of $f_{\rm PBH}$ from CMB observations depending on whether PBHs are assumed to have disk or spherical accretion, and whether PBHs exhibit outflows or not. For a monochromatic population of PBHs of 1000 M$_\odot$, these uncertainties place the upper limit between 10$^{-5}$-10$^{-2}$ \citep{Piga2022}. However, constraints from dwarf galaxy heating also limit $f_{\rm PBH}$ to 10$^{-3}$ at the chosen PBH mass \citep{Lu2021}. Our values were chosen to test the parameter on the very limit of these constraints, and possibly lie above the maximum possible values if the more stringent constraints are to be believed. The lack of growth at $f_{\rm PBH}=10^{-4}$ may therefore mean that 1000 M$_\odot$ PBHs can not act as the seeds to the JWST observed SMBHs. We test the robustness of this result with respect to the mass resolution of DM particles in Appendix \ref{sec:appendix}.

If however, the true constraints allow for $f_{\rm PBH}=10^{-3}$, the results are more optimistic. Starting from $z \sim 25$, significant growth was achieved in 5 (0.06$\%$) PBHs, which were able to sink to the center of DM halos and accrete up to 10$^{4}$-10$^{5}$ M$_\odot$ by the end of the simulation at $z=21$. We show the growth history as a function of redshift in Figure \ref{fig:growth} and contrast it to recent SMBH observations. We guide the reader's eye with the blue shaded region, which represents the Eddington limited growth required to achieve GNZ-11 \citep{Maiolino2023}. Theoretical work instead suggests that massive BHs are more likely to grow through brief ($\sim$kyr) bursts of super-Eddington accretion \citep{Piana2024,Gordon2024,Gordon2025,Lupi2024}, which is supported by a growing number of observations \citep{Du2018,Yue2023,Bhatt2024,Jin2024,Suh2025,Yang2024,Lambrides2024}. The super-Eddington bursts of growth experienced by our PBHs place these IMBHs on/above track to achieve GNZ-11 at 10$^{6}$ M$_\odot$ by $z\sim10$. Our most massive PBH achieved 1.5 $\times 10^5$ M$_\odot$ by the end of the simulation. The red shaded region shows the range of accretion rates required for the most massive PBH to reproduce all of the inlcluded observations. The final PBH mass almost perfectly matches the mass required to achieve UHZ1 \citep{Bogdan2024} with Eddington limited growth by $z\sim10$. All of the collected SMBH observations can be reproduced by assuming a continued growth rate of between 0.3-1 times the Eddington rate, which is consistent with recent PBH growth calculations by \cite{Dayal2025}.

We show density projections of the region containing these 5 growing PBHs in Figure \ref{fig:vertical_subplots}, which gives the reader an idea of the ubiquity of the PBHs in both dense and sparse regions of the Universe at $f_{\rm PBH}=10^{-3}$. The growing PBHs all reside within DM halos, embedded in clusters of other PBHs and dense gas. The evolution of the environment around these PBHs (100 co-moving pc) is shown in Figure \ref{fig:environment}. Before each of the brief periods of growth, the most massive PBH (blue) enters a dense region containing 10$^4$-10$^5$ M$_\odot$ of cold ($\sim$200 K) gas. The presence of cold gas is indicative that the PBH is approaching the center of a halo's gravitational potential; the outskirts of halos ($\sim$kpc) are typically hot (10$^3$-10$^4$ K), with  temperatures dropping towards their center as the abundance of H$_2$ (the primary gas coolant) increases to a gas fraction of $\sim0.1$, allowing the temperature to reach a minimum of a few 100 K on scales of $\sim$1 pc and number densities of 10$^4$ cm$^{-3}$ (e.g. \citealt{Yoshida2007,Glover2008,Greif2011,Prole2023}). Growing PBHs also remained in comparatively low velocity environments throughout the simulation, culminating in bursts of high Bondi-Hoyle rate (10$^{-2}$ to 1 M$_\odot$ yr$^{-1}$) accretion. For comparison, the most massive PBH's nearest neighbor (green) spends the whole simulation in low density regions containing $\sim 10^3$ M$_\odot$ of hot ($>$10$^4$ K) gas, culminating in consistently low ($<$10$^{-8}$ M$_\odot$ yr$^{-1}$ Bondi-Hoyle accretion rates).

Figure \ref{fig:spacing} shows the distribution of distances to the nearest neighboring PBH, which we used as a proxy for the clustering behavior of PBHs. We show the initial PBH distribution at $z\sim 127$ and 4 snapshots spaced equally in terms of cosmic scale factor $a=1/(1+z)$, from $z = 40 - 20$. Over time, the peak of the distribution is shifted to larger physical spatial scales due to the expansion of the Universe. However, by instead looking at the distribution in a co-moving frame (RHS), the peak of the distribution shifts to smaller scales. This implies that the dispersion of PBHs as the Universe expands is slower than the expansion itself, due to the collapse of PBHs into DM halos/PBH clusters. The characteristic scale of this clustering at the end of simulation is $\sim$ 700 pc, consistent with previous findings from purely N-body simulations that PBHs form stable clusters on the scale of 1 kpc \citep{Trashorras2021}, although this probably happens due to the same mechanism that causes DM to form halos on the scale of $\sim$ 1 kpc i.e. the initial velocity and temperature dispersion from $\Lambda$CDM cosmology. Within these clusters, the proportion of PBHs gathered at smaller scales ($<$100 pc) increases towards the end of the simulation, as PBHs sink into the center of DM halos from $z=25$ onward.

No PBH mergers occurred at either of our adopted values of $f_{\rm PBH}$, in contrast to \cite{Tkachev2020}, who also tested this range of $f_{\rm PBH}$ values with N-body simulations. However, they tested lower PBH masses in the range 10-30 M$_\odot$, meaning our 1000 M$_\odot$ were roughly 100 times less numerous to make up the same percentage of DM mass, making PBH mergers in our simulations significantly less likely.

\section{Future work and caveats}
\label{sec:caveats}
The PBHs present at the start of our simulations are the only subgrid objects present throughout the simulation i.e. we do not include star formation, SNe feedback or astrophysical BH formation. The effects of these omissions is unclear; the lack of star/BH formation means that dense gas is freely available for PBHs to accrete, however PBHs would likely merge with these stars/BHs in situations where they would compete to accrete the gas. Likewise, SNe explosions have been shown to lower gas densities in star forming regions, but they have also been shown to push gas towards accreting objects in some cases \citep{Mehta2024}. In future work, we intend to include star/BH formation and SNe feedback. This will have the added benefit of allowing us to asses whether the presence of PBHs shifts star formation to earlier cosmic time and to higher mass halos as in the semi-analytic model of \cite{Liu2022}.

We also have not included PBH accretion feedback in this work, which has the effect of heating the surrounding gas and hindering further accretion. This will also be implemented in future work.

While we varied the fraction of DM that exists as PBHs, we used a monochromatic PBH mass spectrum, with an initial PBH mass of $\sim 1000$ M$_\odot$. The initial PBH mass is another parameter space we will explore in future investigations within the confines of observational constraints. Likewise, different PBH formation models predict different mass functions, such as log-normal and power law mass functions, which will be explored in future work.

Finally, the simulation box in this investigation represents an over-dense region in the Universe. As such, our PBHs had the higher than average change of growing. This further drives our finding that $f_{\rm PBH} = 10^{-4}$ is insufficient to attribute high redshift SMBH seeding to PBHs.

\section{Conclusions}
We have performed the first cosmological simulations to directly include PBHs, testing scenarios where PBHs with initial masses of 1000 M$_\odot$ make up fractions $f_{\rm PBH}=10^{-4}$ and 10$^{-3}$ of all DM. These values are on the upper limit of what current observational constraints suggest is possible. The results are summarized as follows;
\begin{itemize}
  \item At $f_{\rm PBH}=10^{-4}$, no PBHs grew from their initial mass, suggesting that at this fraction of DM, PBHs are not ubiquitous enough to find and sink into the center of DM halos, and hence could not act as the seeds for the JWST observed SMBHs.
  \item At $f_{\rm PBH}=10^{-3}$, 5 PBHs were able to sink into DM halos and grow up to 10$^{4}$-10$^{5}$ M$_\odot$ by $z=20$. This shows that the $f_{\rm PBH}=10^{-4}-10^{-3}$ boundary marks the threshold of where 1000 M$_\odot$ PBHs can act as the seeds for SMBHs. However, these values of $f_{\rm PBH}$ are within the uncertainty regions of the maximum allowed PBH abundances constrained by CMB and dwarf galaxy observations. 1000 M$_\odot$ PBHs may therefore not be capable of acting as SMBH within the allowed parameter space once constraints become more certain.
  \item Our PBHs grew when they entered regions of cold, dense gas. While BHs resulting from the remnants of Population III stars usually already form in such regions, the random conversion of DM particles into PBH at high redshift makes it rare for PBHs to find regions suitable for growth. From our 8726 PBHs, only 0.06$\%$ were able to interact with such environments and grow.
\end{itemize}
This work is the first step towards self-consistently modeling PBHs in cosmological simulations. The next steps are to perform simulations with PBHs and ongoing star/BH formation, including SNe/accretion feedback, and exploring more of the M$_{\rm PBH}$-$f_{\rm PBH}$ parameter space.
\label{sec:conclusions}

\section*{Acknowledgements}
LP and JR acknowledge support from the Irish Research Council Laureate programme under grant number IRCLA/2022/1165. JR also acknowledges support from the Royal Society and Science Foundation Ireland under grant number URF\textbackslash R1\textbackslash 191132. 
\ \
The simulations were performed on the Luxembourg national supercomputer MeluXina.
The authors gratefully acknowledge the LuxProvide teams for their expert support.
\ \ 
The authors wish to acknowledge the Irish Centre for High-End Computing (ICHEC) for the provision of computational facilities and support.
\ \
The authors acknowledge the EuroHPC Joint Undertaking for awarding this project access to the EuroHPC supercomputer Karolina, hosted by IT4Innovations through a EuroHPC Regular Access call (EHPC-REG-2023R03-103).

\newpage
\bibliographystyle{aasjournal}
\bibliography{references}

\appendix
\section{Dark matter mass resolution}
\label{sec:appendix}
\begin{figure*}
\centering
          \includegraphics[width=1\linewidth]{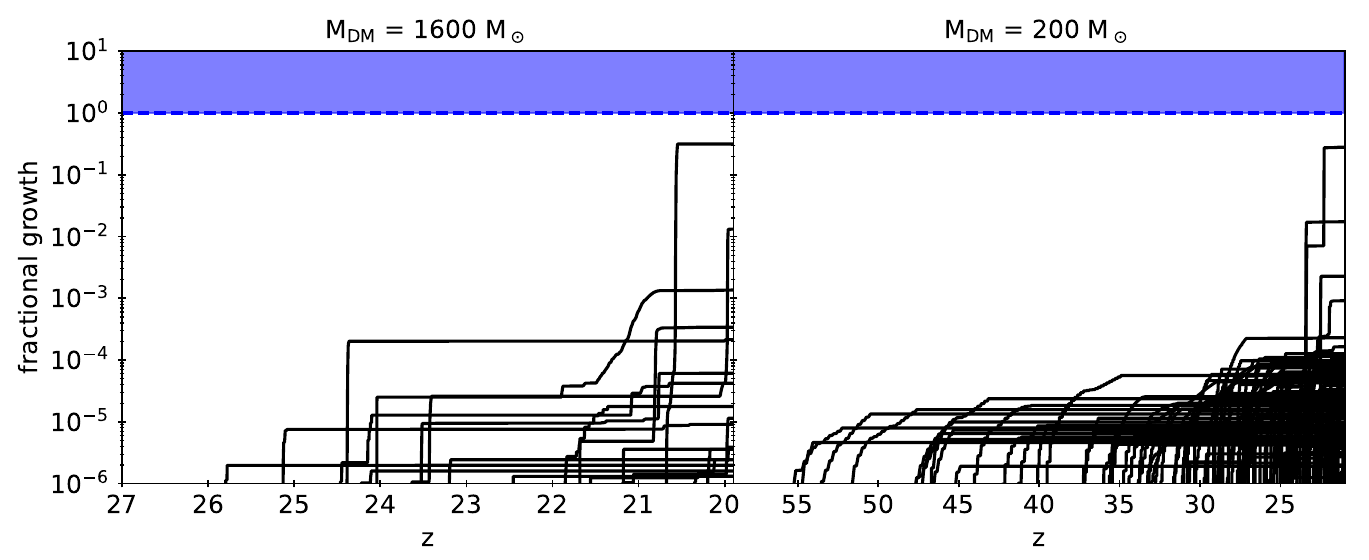}
    \caption{Fractional growth of PBHs at $f_{\rm PBH}$ = $10^{-4}$ as a function of redshift, compared for a DM mass resolution of 1600 and 200 M$_\odot$. The blue shaded region indicates a doubling of the initial PBH mass.}
    \label{fig:resolution}
\end{figure*}

The ability of PBHs to embed themselves within halos may be affected by the mass resolution of DM particles. For example, if DM particles are lower mass than PBHs, they may collapse around PBHs as a result of the PBH's gravity - PBHs would therefore not need to locate halos, as the halo would form around it naturally. The validity of our finding that 1000 M$_\odot$ PBHs at $f_{\rm PBH}$ = $10^{-4}$ do not experience significant growth by $z = 20$ is therefore dependent on whether this holds when the DM particles are more highly resolved. To investigate this, we repeat the $f_{\rm PBH}$ = $10^{-4}$ simulation with 15 levels of refinement, giving a DM particle mass of 200 M$_\odot$, 8 times smaller than the fiducial simulations. To keep the same PBH mass, we converted clusters of 8 DM particles into single PBHs with the combined cluster mass. After randomly sampling DM particles for conversion, we selected the 7 nearest neighboring DM particles to be absorbed into the PBH. The conversion process resulted in a set-up with 791 PBHs, 68914672 top grid DM particles and gas cells, giving $f_{\rm PBH}$ = 0.918$\times 10^{-4}$. We ran the simulation down to $z=21$ to match the fiducial run. We show the PBH growth as a fraction of their initial mass for both resolutions at $f_{\rm PBH}$ = $10^{-4}$ in Figure \ref{fig:resolution}. The result was that no PBHs were able to grow significantly despite the introduction of DM-PBH mass disparity, reinforcing our finding that PBHs do not act as the seeds to the JWST observed, high redshift SMBHs. We do find that improved DM mass resolution leads to earlier PBH growth, beginning at $z=55$ instead of 26 in the low resolution case. This is expected as the higher resolution allows smaller sub-halos to form, increasing the number of viable accretion sources and improving the likelihood of PBHs interacting with dense gas. However, the accretion onto the most massive PBH remained unchanged.

\end{document}